\documentclass[10 pt,a4 paper, twoside]{article}
\usepackage[utf8]{inputenc}
\usepackage{amsmath}
 \usepackage{graphicx}
 \usepackage{grffile}

\title{Another Kerr interior solution}
\author{ Fatemeh Zahra Majidi\thanks{fzmajidi@ut.ac.ir}}
\date{\textsl{\small{Department of Physics, University of Tehran, Tehran 14395-547, Iran}}}
 
\begin{document}
 
\begin{titlepage}
\maketitle
\begin{center}
\textbf{Abstract}
\end{center}
\vspace{0.5 cm}
A stationary axially symmetric solution describing a rotating anisotropic source for Einstein Field Equations(EFE) is proposed which matches to the exterior Kerr metric. The anisotropic source satisfies all energy conditions - weak, strong, and dominant energy condition - for a wide range of metric's free parameters values. The resultant energy-momentum tensor components, and consequently energy density and pressure profiles, redshift function and angular velocity are singularity free and behave as expected. As rotation parameter goes to zero a spherical source consisting of normal and exotic matter is retrieved for the exterior Schwarzschild solution; the corresponding redshift function, however, is well-behaved and positive definite. This is the first solution of its kind.
\section{Introduction}

Ever since 1963 that R. P. Kerr presented his exact solution for EFE \cite{kerr}, numerous attempts for finding a physically reasonable interior solution serving as its source have been made [2-43]; however, all of them proved to be unsuccessful so far. According to Krasinski's review \cite{krasinski} these attempts can be categorized into four groups. As an introduction to this problem, an up-to-date summary of Krasinski's review - which covers papers until 1975 - is presented here, allocated to the papers specifically devoted to Kerr or Kerr-Newman interior solutions.

\vspace{0.2cm}

\textbf{1. Identifying possible sources by investigation of singularities and physical interpretation of parameters of the Kerr metric.}

Papers in this category [8-16] are the very first attempts in order to show what the basic properties of a plausible source should be. \cite{bp}, \cite{leau}, \cite{md},\cite{cohen}, and \cite{lt}, as Krasinski correctly concluded, provided a preliminary picture of the source, though a very vague one ("something that rotates" was the message). Newman and Janis's source \cite{nj1} is a rotating ring of uniformly distributed mass and charge - a source for the Kerr-Newman metric rather than Kerr - placed inside the exterior Kerr metric's ring singularity. The interior metric coefficients, however, suffer a non-uniqueness of a branch-point type problem (multivalued solutions are generated in fact from their source). Kerr himself proposed that it is probably a disk that should span the ring singularity. This remark was a turning point, for after publication of \cite{nj1} relativistic disks became the favorite and most investigated sources for the exterior Kerr metric.    \end{titlepage} 

\textbf{2. Eliminating some types of sources due to contradictions or inconsistencies to which they lead.}

None of the papers cited by Krasinski in this group [16-25] were successful in ruling out a specific type of source for the Kerr metric, and to the date no paper successful in this regard has been published. 

Boyer \cite{by1} \cite{by2} studied the properties of boundary surfaces of rigidly and non-rigidly rotating perfect fluid sources. The result is highly arbitrary in metric components, and therefore not practically useful. Hernandez \cite{h1} \cite{h2} tried to rule out perfect fluid as a physically reasonable source for the exterior Kerr metric, but his basic assumptions proved to be wrong \cite{krasinski}. Herlt \cite{herlt} claimed a rigidly rotating perfect fluid can not be a  physically meaningful source for the exterior Kerr metric, but again like the previous case, his result was based on mistakes in analysis \cite{roos}. Abramowicz et. al. also tried to put bounds on fluid sources \cite{abm} or rule out pressure-less disks as sources of the Kerr metric \cite{abm2} but both of their papers proved to be unsuccessful and not useful in practice.

\vspace{0.2cm}

\textbf{3. Constructing physically acceptable configurations matched to the Kerr metric only approximately.}

Papers belonging to this group are very well-known in the Kerr metric's interior solutions literature [26-31].  

Cohen \cite{coh} showed a rotating spherical shell of zero thickness can be matched to Kerr metric up to linear terms in the angular velocity, and concluded a full perfect fluid sphere - as well as many other possible spherical sources - can be an approximate source for the Kerr metric in this formalism. de la Cruz and Israel \cite{ci} confirmed Cohen's results to first order and found out in the second order all the properties of the shell will be lost unless it is placed right on the Schwarzschild sphere. McCrea \cite{mc} showed that up to $k^2$ ($k$ as a small parameter) the exterior field of a rigidly rotating material sphere is Kerr metric. Florides \cite{fl} showed up to $k^5$ the Kerr metric can be matched to a rotating sphere of anisotropic fluid; up to $k^3$ its rotation is rigid and in higher orders it becomes nonrigid. 

It would be noteworthy to mention the papers [33-36] in which rotating sources made of various fluids have been studied, although the main aim of these papers was not constructing sources for the exterior Kerr metric - but more general stationary exterior metrics which could reduce to the Kerr metric as one of the possible cases. One can find valuable information on to what order Kerr metric can be matched to the considered rotating sources in these papers - for some sources in the slowly rotating limit as well. The resultant energy density and pressure profiles, however, experience infinities in central region. Also, a classification of these sources according to various Petrov types are given in \cite{pet}.

\vspace{0.2cm}

\textbf{ 4. Constructing odd configurations with unphysical properties matched to the Kerr metric exactly.}

Papers in this group [2][38-43] are the earliest attempts for satisfying junction conditions \cite{poi}. Amongst the papers in this category, Gurses and Gursey \cite{gg} and Collas and Lawrence \cite{cc} found anisotropic sources with at least one negative pressure profile. The proposed solution in this paper and another recent work \cite{her} - also describing an anisotropic source - both have this feature. It is worth mentioning that later Collas \cite{col} checked the proposed solution in \cite{cc} and found out the corresponding redshift becomes discontinuous in central region.

Keres \cite{is1} and Israel \cite{is2} found a disk spanned by a ring of radius $a$; the interior of the disk has negative mass whose density diverges to $- \infty$ as approaching the circumferential ring, and since the ring has infinite positive mass the net mass remains finite. Israel has mentioned in his paper \cite{is2} that maybe this solution can be interpreted as a gravitational model for electron (this argument was later followed by Burinskii's microgeon model made of gravitationally bound electromagnetic fields in a series of papers started by \cite{bur}).

Hogan \cite{hog} obtained an anisotropic source enveloped in a crust of zero thickness with finite surface matter density. This solution is also investigated by Collas \cite{col}: the redshift function corresponding to this solution becomes complex in central region. Hamity \cite{ham} found a rigidly rotating disk with regular interior and singular edge which has zero energy density and isotropic pressure that diverges to $\infty$ on the edge.

More recent works belonging to this group try to satisfy Dormois - Israel \cite{dorm} \cite{iss}  boundary conditions \cite{drake} \cite{vi}; the solution proposed by Drake and Torulla \cite{drake} matches smoothly to the exterior Kerr metric but its pressure profile diverges in central region. Viaggiu's solution \cite{vi} also satisfies Dormois - Israel matching conditions and reduces to Schwarzschild incompressible fluid \cite{sch} in the spherically symmetric limit, but yet it is incomplete as there is no ansatz satisfying energy conditions. \cite{drake} and \cite{vi} are both based on Newman and Janis Algorithm \cite{nj1}. Kyriakopoulos \cite{ky} has also suggested a family of EFE solutions serving as interior solutions for the exterior Kerr metric that are all singular.

\vspace{0.2cm}

Now it is necessary to add a fifth group to Krasinski's classification:

\vspace{0.2cm}

\textbf{5. Sources made of physically acceptable configuration that match to the Kerr metric exactly, but experience ultra-relativistic limits in central region.}  

 The best-known solution in this group is proposed by Neugebauer and Meinel \cite{meinel}: a rigidly rotating disk of dust which approaches the extreme Kerr solution ($J=M^2$) under special conditions. The disk, however, exhibits a discontinuity which was interpreted by the authors as a phase transition from normal matter to the black hole state. Another solution of this type can be found in Pichon and Lynden-Bell's paper \cite{pl} which generates anisotropic disks that match the exterior Kerr metric smoothly and analytic expressions were found for their energy density and pressure profiles. The only problem this solution has is again the divergence of central redshift function. In this regard, both solutions experience ultra-relativistic limits in the center of the generated disks. 

\vspace{0.2cm}

There are also two important papers which can't be placed in the above categories:

1. Bicak and Ledvinka \cite{led} proposed relativistic disks as sources of the Kerr-Newman exterior metric: these disks satisfy weak energy condition (WEC) and strong energy condition (SEC) but become highly relativistic in central region. These disks can not be matched smoothly to the exterior Kerr metric, but they are not approximate solutions either; although discontinuous at the boundary surface, they can be matched to the exterior Kerr metric via a thin shell \cite{poi} with a finite, physically reasonable stress-energy tensor. The suggested solution in this paper has this feature as well. The spin zero limit of the disks has not been checked.

2. Another impressive solution published recently \cite{her} - which can not be categorized yet as its corresponding redshift function has not been studied - generates anisotropic oblate sources which match the exterior Kerr metric smoothly; in its spherically symmetric limit the Schwarzschild interior metric \cite{sch} is retrieved and for a wide  range of parameters the solution satisfies SEC. As mentioned above and proposed earlier by Collas \cite{col}, checking the redshift function - aside from the energy-momentum tensor components - can be the very first test for verifying the validity of the solution and according to private communication between the author and L. Herrera the investigation of redshift function is still in progress. If the resultant redshift shows no irregularity, \cite{her} will be the very first interior solution for the exterior Kerr metric.

\vspace{0.2cm}

As regards the above review, the major problems of the so-far-suggested interior solutions can be identified now. Based on these problems, a list of essential criteria that every well-behaved, continuous interior solution for the exterior Kerr metric should adhere to can be extracted. A physically, reasonable solution should:  

1. \hspace{0.2cm}Match the exterior Kerr metric smoothly or on a thin shell;

2. Favor at least one of the WEC, SEC, or dominant energy condition (DEC);

3. In addition to the energy-momentum tensor components, the interior solution itself should be singularity free everywhere;

4. The resultant redshift function and angular velocity should be well-behaved and positive definite;

5.\hspace{0.2cm} Energy density and pressure profiles should monotonically decrease from the center of the body toward its surface.

\vspace{0.2cm}

The proposed interior metric in this paper is the first attempt for suggesting a practical and flexible source in order to satisfy all the energy conditions, SEC, WEC and DEC, and to have finite and positive definite redshift and angular velocity; that according to the above review of the so-far-suggested interior solutions is a novel result. The spherically symmetric limit of this solution is made of two fluids - one that satisfies SEC and the other which violates energy conditions - but as expected can be matched to the exterior Schwarzschild metric. A discussion on whether this solution is physically allowable or not is presented in the discussion section. 

This paper is structured as follows: in section 2, a stationary axisymmetric anisotropic source is proposed for the Kerr metric - as an ansatz - which matches the Kerr metric on a thin shell. All the resultant physical variables are singularity free and behave as one expects from a rotating source; i.e. the angular velocity increases from zero as one moves outward from the center of the body, redshift function remains finite and positive everywhere, energy density and pressure profiles monotonically decrease from the center of the body toward its surface. In section 3, the spherically symmetric limit of the suggested solution is studied. The spherical source consists of two different fluids and matches the exterior Schwarzschild metric on a thin shell. The resultant redshift function is well-behaved and positive definite, although one of the source fluids is exotic.

\section{The interior metric}

The seed metric for every stationary axially symmetric metric can be written in the following form \cite{col}:

\[ ds^2= - e^{2\nu (r,\theta)} \hspace{0.1cm} dt^2 + e^{2\psi(r,\theta)} (d\phi - \omega(r,\theta) \hspace{0.1cm} dt)^2 + e^{2 \mu _{1} (r,\theta)} \hspace{0.1cm} dr^2 + e^{2 \mu_{2} (r,\theta)} \hspace{0.1cm} d\theta ^2  \hspace{0.1cm},\]

in which $\omega$ is the angular velocity of the locally non-rotating frame (LNRF) observed from infinity and $\nu$ can be considered as the general-relativistic gravitational potential. The proposed metric is a stationary axisymmetric interior solution for EFE in Boyer-Lindquist \cite{bl} coordinates :

\[ ds^2 =- [\frac{\Sigma \Delta}{ B}-\frac{ r_{s} a^2 r^2 (1 - (5+ cos \theta)^2)}{B \Sigma}] dt^2-\frac{ 2 a r_{s} r (1 - (5+ cos \theta)^2)}{\Sigma}  \hspace{0.1cm} d\phi dt  \]
\begin{equation}
+ \frac{\Sigma}{\Delta}  \hspace{0.1cm} dr^2+ \Sigma  \hspace{0.1cm} d\theta ^2 + \frac{ B (1 - (5+ cos \theta)^2)}{\Sigma}  \hspace{0.1cm} d\phi ^2 ,
\end{equation}

where:

\[ \Sigma = r^2 _{g} - a^2 (5 + cos\theta)^2 \hspace{0.1cm}, \]

\[\Delta = \frac{ r_{g} ^3 - r (- a^2 - r_{s} r_{g})}{r_{g}} \hspace{0.1cm} , \]

\[ B = (r^2 _{g} - a^2)^2 + a^2 \Delta (1 - (5+ cos \theta)^2) \hspace{0.1cm}, \]

and $r_{s} = 2 G M/ c^2 $ , $r_{g}$ is the rotating body's radius and $a=J/M$ is the rotation parameter. According to the above metric, it is easy to find the angular velocity and redshift function(for photons emitted in a LNRF) \cite{col}:

\begin{equation}
\omega = \frac{ a r_{s} r}{ B} \hspace{0.1cm},
\end{equation}

\begin{equation}
Z= \sqrt{\frac{B}{\Sigma \Delta}} -1 \hspace{0.1cm} .
\end{equation}

In the same coordinates - for convenience in comparison - one can write the exterior Kerr metric as follows:

 \[ ds^2 =- [\frac{\Sigma \Delta}{ B}+\frac{ r_{s} a^2 r^2 sin^2 \theta}{B \Sigma}] dt^2-\frac{ 2 a r_{s} r  sin^2 \theta}{\Sigma} \hspace{0.1cm} d\phi dt + \frac{\Sigma}{\Delta}  \hspace{0.1cm} dr^2 \hspace{0.1cm}   \]
\begin{equation}
+ \Sigma  \hspace{0.1cm} d\theta ^2 + \frac{ B  sin^2 \theta}{\Sigma}  \hspace{0.1cm} d\phi ^2 ,
\end{equation}

where:

\[ \Sigma = r^2 + a^2  cos^2 \theta \hspace{0.1cm}, \]

\[\Delta = r^2 - r_{s} r + a^2 \hspace{0.1cm} , \]

\[ B = (r^2 + a^2)^2 - a^2 \Delta \hspace{0.1cm} sin^2 \theta \hspace{0.1cm}, \]

in which $r_{s} = 2 G M/ c^2 $  and $a=J/M$ is the rotation parameter. Metric (1) can be matched to the exterior Kerr metric on a thin shell \cite{poi} with curious features. This thin shell has jumps in the rotation parameter and $cos \theta$ function:      at the boundary surface ($ r= r_{g}$) by substituting the exterior $(cos\theta)^2$ with $(5 + cos\theta)^2$ and $a^2$ with $- a^2$ the interior solution will be retrieved. The existence of this thin shell in turn implies that $T_{\mu \nu}$ is non-zero at the surface layer and the two metrics can not be matched smoothly because of the change of sign in $a^2$. 

$(5 + cos\theta)^2$ is achieved through checking the resultant redshift functions for various values of $(n + cos\theta)^2$. For $n = 0,1$ , $(cos\theta)^2$ and $(1 + cos\theta)^2$ correspond to singular interior solutions at $\theta = \pi/2$ and $ \theta = 0,\pi$ respectively. For $n = even \hspace{0.1cm} numbers$ the resultant redshift - although continuous - generates both positive and negative values, but for $n = odd \hspace{0.1cm} numbers $ and $ n \geq 5$ - as desired - the redshift is positive definite. In this paper, just the $n =5$ cases is considered; for higher odd numbers the results are all the same and only different in energy density, pressure profiles, angular velocity, and redshift function values. Larger values of $n$ result in lower values for the mentioned functions. From this aspect, the suggested solution can generate numerous physically allowable rotating, anisotropic sources for the exterior Kerr metric - as real planets and neutron stars.

The change of variables should have a meaningful physical interpretation, but at the moment the suggested solution can be considered as a workable guess for EFE which matches the exterior Kerr metric. Certainly, as it is evident, Boyer-Lindquist coordinates is not the best coordinate system for describing metric (1).

For obtaining the energy density and pressure profiles of the source, the following eigenvalue equation should be solved:

\begin{equation}
|T_{\mu \nu} - \lambda_{i} g_{\mu \nu}| = 0  \hspace{0.1cm},
\end{equation}

in which $-\lambda_{0}$ indicates energy density and $\lambda_{i}$ are pressure profiles to the first order of $a$. Solving (5) for metric (1) results in \cite{vi} :

\[
\begin{vmatrix}
T_{tt} - \lambda g_{tt} & 0 & 0 & T_{t\phi}-\lambda g_{t\phi} \\
0 & T_{rr} -\lambda g_{rr} & T_{r\theta} & 0 \\
0 & T_{r\theta} & T_{\theta \theta} - \lambda g_{\theta \theta} & 0 \\
T_{t \phi} - \lambda g_{t\phi} & 0 & 0 & T_{\phi \phi} - \lambda g_{\phi \phi}
\end{vmatrix}
= 0
\]

and the eigenvalues are given by:

\[ \lambda_{0,1} = \frac{P\pm \sqrt{P^2 -4(g_{tt} g_{\phi \phi} - g^2 _{t \phi})(T_{tt} T_{\phi \phi} - T^2 _{t\phi})}}{2[g_{tt} g_{\phi \phi} - g^2 _{t\phi}]} \hspace{0.1 cm} , \]

\[ P=(T_{tt} g_{\phi \phi} + g_{tt} T_{\phi \phi} - 2 T_{t\phi} g_{t\phi}) \hspace{0.1 cm}, \]

\[ \lambda_{2,3} = \frac{Q\pm\sqrt{Q^2 - 4g_{rr} g_{\theta\theta} (T_{rr} T_{\theta \theta} - T^2 _{r\theta})}}{2g_{rr} g_{\theta\theta}} \hspace{0.1 cm} , \]
\[ Q=(T_{rr} g_{\theta\theta} + g_{rr} T_{\theta \theta}) \hspace{0.1 cm} . \]

Regarding energy conditions, for the weak energy condition eigenvalues should satisfy:

\[ -\lambda_{0} \geq 0 \hspace{0.2cm},\hspace{0.2cm} -\lambda_{0} + \lambda_{i} \geq 0 \hspace{0.1 cm} ; \]

for the strong energy condition :

\[ -\lambda_{0} + \sum_{i} \lambda_{i} \geq 0 \hspace{0.2cm},\hspace{0.2cm} -\lambda_{0} + \lambda_{i} \geq 0 \hspace{0.1 cm} ; \]

and for the dominant energy condition :

\[ -\lambda_{0} \geq 0 \hspace{0.2cm},\hspace{0.2cm} \lambda_{0} \leq \lambda_{i} \leq - \lambda_{0} \hspace{0.1 cm} . \]

Determining an equation of state is a rather complicated task, so instead a practical example is studied. The free parameters can be fixed according to the available information on planet earth(in SI units): $M = 5.972 \times 10^{24} \hspace{0.1cm} kg$ , $a = J/M \simeq 10^{16} \hspace{0.1cm}m^2/s $ , $r_{g} \simeq 10 ^6 \hspace{0.1cm} m$ , and therefore $r_{s} \simeq 0.001 \hspace{0.1cm} N s^2/kg$ . Due to change of variables, the usual $0 \leq \theta \leq \Pi$ range changes to $0 \leq \theta \leq 2 \Pi$. The energy density, pressure profiles, redshift function, and angular velocity are plotted according to the fixed free parameters in Figure 1. All these physical variables are well-behaved and singularity free.

\begin{frame}

\centering
\includegraphics[width=5cm,height=4cm]{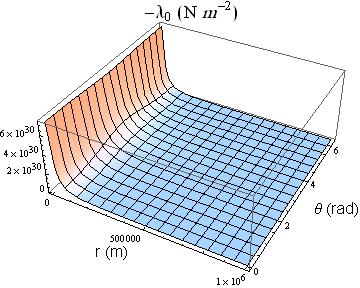} \   \includegraphics[width=5cm,height=4cm]{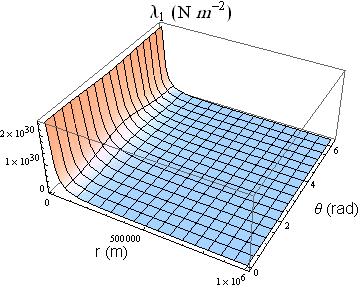} 

\centering
{(a)} \hspace{4.5cm} {(b)}
\vspace{0.5cm}

\centering
\includegraphics[width=5cm,height=4cm]{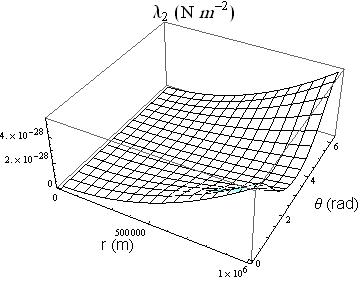} \    \includegraphics[width=5cm,height=4cm]{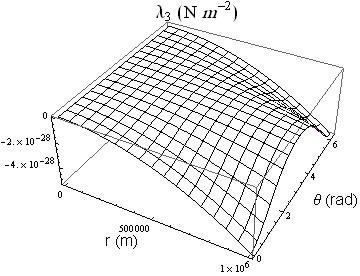} 

\centering
{(c)} \hspace{4.5cm} {(d)}
\vspace{0.5cm}

\centering
\includegraphics[width=5cm,height=4cm]{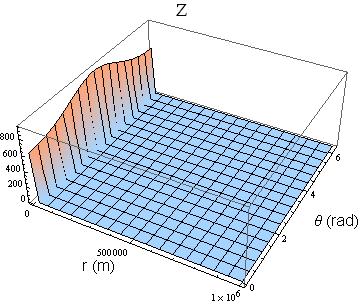} \    \includegraphics[width=5cm,height=4cm]{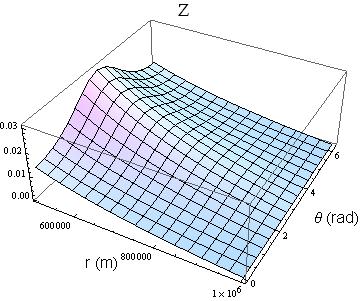}  

\centering
{(e)} \hspace{4.5cm} {(f)}

\centering
\includegraphics[width=5cm,height=4cm]{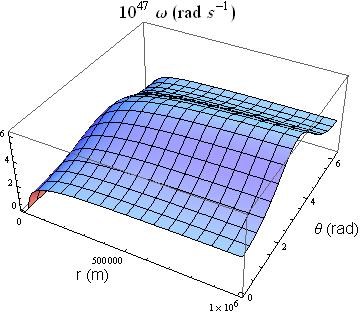} \   

\centering
{(g)}

\end{frame}
 \vspace{ 0.5cm}

Figure 1 : The energy density (a), pressure profiles (b)(c)(d), angular velocity (g) and redshift function (e)(f) diagrams for a toy model with the same mass, radius, and angular momentum as planet earth. All diagrams are singularity free.
 \vspace{ 0.5cm}

Now it is  easy to check the energy conditions through graphs plotted in Figure 2; all the SEC, WEC, and DEC are favored.

\begin{frame}

\centering
\includegraphics[width=5cm,height=4cm]{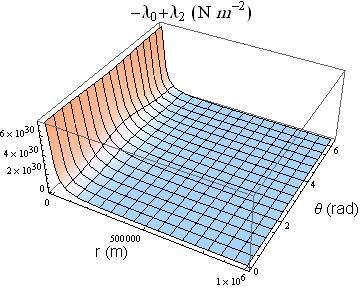} \    \includegraphics[width=5cm,height=4cm]{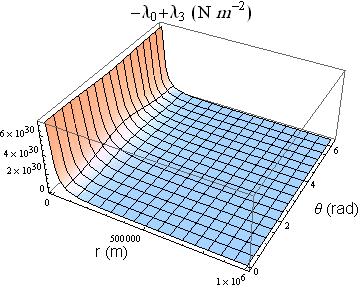} 

\centering
\includegraphics[width=5cm,height=4cm]{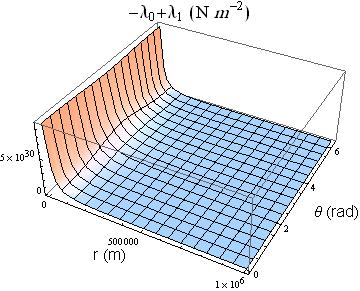} \    \includegraphics[width=5cm,height=4cm]{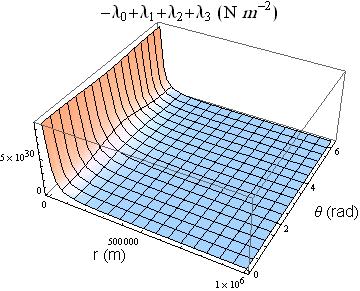}

\end{frame}
 \vspace{ 0.5cm}

Figure 2 : Graphs for checking WEC ($ -\lambda_{0} + \lambda_{i} \geq 0 $) and SEC ($-\lambda_{0} + \sum_{i}   \lambda_{i} \geq 0$ and simultaneously $-\lambda_{0} + \lambda_{i} \geq 0$). As regards the energy density is positive definite and $\lambda_{1}$, $\lambda_{2}$, and $\lambda_{3}$ are all smaller in amount compared to the energy density, strong,weak and dominant energy conditions are satisfied.
 \vspace{ 0.5cm}

Table 1 indicates a sample with the same mass as planet earth but different radius and rotation parameter. The aim of this table is to exhibit in what range of parameter values energy conditions are favored.

\section{The spherically symmetric limit}

As rotation parameter goes to zero, a spherically symmetric metric is achieved:
\begin{equation}
 ds^2 =- [\frac{r^2_{g}- r_{s} r}{ r^2_{g}}] dt^2 + \frac{r^2_{g}}{r^2_{g}- r_{s} r}  \hspace{0.1cm} dr^2 + r^2_{g}  \hspace{0.1cm} d\theta ^2 + r^2 (1 - (5+ cos \theta)^2)  \hspace{0.1cm} d\phi ^2 , 
\end{equation}

that matches the exterior Schwarzschild metric: 

\begin{equation}
ds^2= -(1-\frac{r_{s}}{r}) dt^2 + (1-\frac{r_{s}}{r})^{-1} dr^2 + r^2 (d\theta ^2 + sin^2\theta d\phi ^2)  \hspace{0.1cm} ,
\end{equation}

through a thin shell on which the exterior $(cos\theta)^2$ turns to  $(5 + cos\theta)^2$ in spherical coordinates. Metric (6) describes a fluid which experiences two distinct regimes: $r_{g}\leq r_{s}$ and $r_{g}> r_{s}$ which is evident from the metric structure, redshift function, and energy-momentum tensor components. The corresponding redshift function is:
\begin{equation}
Z= \frac{\sqrt{r^2_{g} - r_{s} r}}{r_{g}} -1 \hspace{0.1cm}.
\end{equation}

This function generates negative values as well as positive values for $r_{g}\leq r_{s}$ regime, hence it will not be considered in this paper; although the resultant energy density and pressure profile will be finite everywhere. As the metric is diagonal the energy density and pressure profiles can be read directly from energy momentum tensor components:

\begin{equation}
T_{tt} = \rho / c^2 = \frac{cos \theta (r^3 _{g} - r_{g} r_{s} r)}{r^5 _{g} (5+cos \theta)} \hspace{0.1cm},
\end{equation}

\begin{equation}
T_{rr} = P_{r} = \frac{r_{g} cos\theta}{(r^3 _{g} - r_{g} r_{s} r) (5+cos \theta)} \hspace{0.1cm},
\end{equation}

and the $T_{\theta \theta}$ and $T_{\phi \phi}$ components are both zero. It is easy to write an equation of state for this spherical source:

\begin{equation}
\rho / c^2 = \frac{(r^3_{g} - r_{g} r_{s} r)^2}{r^6 _{g}} P_{r} \hspace{0.1cm}.
\end{equation}

Figure 3 indicates the energy density, radial pressure, and redshift function for a spherical planet with $r_{s}=10^{-6} \hspace{0.1cm} N s^2/kg$ and $r_{g}=10^{6} \hspace{0.1cm} m$. 

\begin{frame}

\centering
\includegraphics[width=5cm,height=4cm]{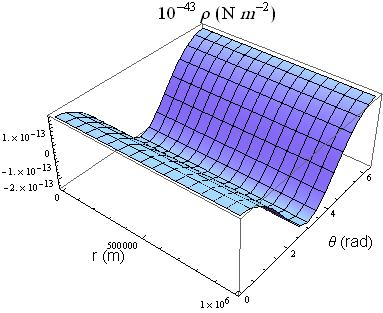} \    \includegraphics[width=5cm,height=4cm]{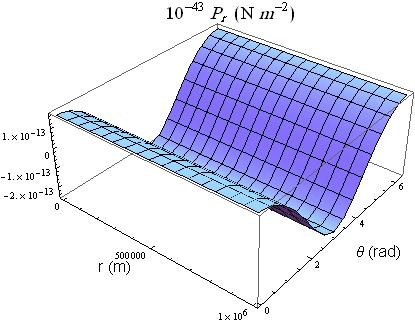} 

\centering
{(a)} \hspace{4.5cm} {(b)}

\centering
\includegraphics[width=4cm,height=3cm]{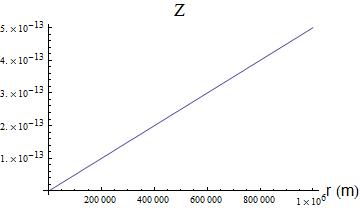}\  \hspace{0.5cm}  \includegraphics[width=5cm,height=3cm]{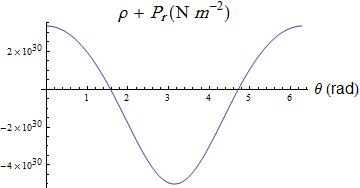}

\centering
{(c)}  \hspace{4.5cm} {(d)}

\end{frame}
 \vspace{ 0.5cm}

Figure 3 : The energy density (a), radial pressure (b),  and redshift function (c) for a spherical planet with $r_{g} = 10 ^{6} \hspace{0.1cm} m$ and $r_{s} = 10^{-6} \hspace{0.1cm} N s^2/kg$ . 

\vspace{ 0.5cm}

According to Figure 3 (d), it is evident that SEC is favored only for half of the source and the other half violates all energy conditions. It is interesting that although the source consists of exotic matter, the redshift function is well-behaved and positive definite. Therefore, the spherical body does not experience an ultra-relativistic limit or a phase transition from normal matter to black hole state as explained in the introduction. 

\section{Discussion}

Table 2 is a check list allocated to the proposed stationary, anisotropic solution and its spherically symmetric limit properties.

\begin{table}[ht]
\centering
\resizebox{\textwidth}{!}{\begin{tabular}{ l | c|c  r }
  \hline
\rule{0pt}{3ex}   			
 \hspace{0.2cm}Anisotropic, Rotating Source & Properties & Static Source \\
\hline
\rule{0pt}{3ex}   
Positive definite and continuous  & Redshift &   Positive definite and continuous  \\
 \hline
\rule{0pt}{3ex}   
Satisfies SEC, WEC, and DEC  & Energy Conditions & Half of the sphere satisfies SEC  \\
 \hline
\rule{0pt}{3ex}   
Positive definite and continuous  & Angular velocity & - \\
 \hline
\rule{0pt}{3ex}   
\hspace{1.9cm} None & Singularities & None\\
\hline
 
\end{tabular}}
\end{table} 

 Table 2: Properties of the proposed stationary, anisotropic source and its spin zero limit.
 \vspace{ 0.3cm}

The above table shows that a spherical source for the exterior Schwarzschild metric containing exotic matter - partially - has turned into normal matter through rotation. Some may argue that a valid, physically allowable stationary solution for the exterior Kerr metric must reduce to a well-known source for Schwarzschild metric in spin zero limit; since Kerr metric reduces to Schwarzschild metric as rotation parameter goes to zero. It would be very desirable, indeed, like the cases studied in \cite{vi} and \cite{her}; but it is not necessary. An interior solution for Kerr metric satisfying all energy conditions can reduce to a source for Schwarzschild metric in spin zero limit, with no guarantee that this generated space-time may also satisfy all or even one of the energy conditions. A good example may be the anisotropic source proposed in \cite{her}; this source only satisfies SEC, but in the spherically symmetric limit turns into Schwarzschild incompressible fluid which satisfies all energy conditions - SEC, WEC, and DEC. The authors of \cite{her} have assumed what conditions the incompressible fluid will impose on their stationary solution at first place and based on these conditions have constructed their source. In fact, they have followed intuition rather than a clear mathematical procedure in this regard. Hence, it seems there is no direct correspondence between the stationary and static sources. The condition that a meaningful, physically allowable stationary source should reduce to a plausible static source in spin zero limit appears to be more of an assumption than a  necessity or criterion. 

So far many methods have been proposed for generating stationary sources \cite{pl} [33-36] \cite{dorm} \cite{iss} [53-58], but none of these methods mathematically entail that in the spherically symmetric limit the energy conditions must be satisfied as well; although by intuition this assumption seems true, it is not a mathematical fact yet. For example, the highlighted papers on this problem such as \cite{meinel}, \cite{pl}, and \cite{led} have not pointed out this issue unlike \cite{drake}, \cite{vi}, and \cite{her}. Therefore a broad range of possible exact solutions should be studied to confirm Kerr interior solutions must reduce to interior Schwarzschild solutions, satisfying energy conditions - it is worth mentioning that even the known well-behaved, continuous solutions for the Schwarzschild exterior metric are few in number \cite{ste}. For example, yet, it is not known an anisotropic, perfect fluid or dust source for the exterior Kerr metric must correspond to which of incompressible Schwarzschild fluid, perfect fluid or dust sources for the exterior Schwarzschild solution. For the studied case in this paper, as an example, it is evident that the eigenvalue equation $|T_{\mu \nu} - \lambda_{i} g_{\mu \nu}| = 0$ is useless for realizing what source the anisotropic source - as the most probable source for the Kerr metric \cite{vi} - must reduce to in spin zero limit. For a Schwarzschild interior solution which does not contain off-diagonal Einstein tensor components, writing such an eigenvalue equation is meaningless. Hence this issue raises the question that whether the static sources generated by valid stationary interior solutions should also satisfy energy conditions; or whether there is a rule with mathematical basis implying the corresponding Schwarzschild interior solutions must inherit the energy conditions the stationary solution satisfies; or amongst the energy conditions the stationary solution satisfies which ones the static case must favor? These are the open, interesting questions which may lead to the construction of new methods for suggesting new sources for Kerr metric or other exterior stationary space-times.

\section{Conclusion}

The proposed stationary axisymmetric solution which describes a rotating anisotropic source matches the Kerr metric on a thin shell and as a result, the energy-momentum tensor is non-zero at the boundary surface. The corresponding redshift function and angular velocity are positive definite and well-behaved. The resultant energy density and pressure profiles favor SEC, WEC, and DEC simultaneously; hence the suggested solution can successfully describe a rotating stationary axisymmetric body. The energy-momentum tensor, redshift function, and angular velocity are all singularity free. The solution can be considered as a very flexible model for rotating planets and neutron stars, and therefore useful for data-fitting purposes. In the spherically symmetric limit, the interior metric matches the exterior Schwarzschild metric on a thin shell and again the energy-momentum tensor is non-zero at the surface layer. As an interior solution, half of the spherical source favors SEC and the other half violates all energy conditions. The corresponding redshift, however, is continuous and positive definite, hence, the exotic matter can not be interpreted as a phase transition from normal matter to the black hole state; i.e. the source does not experience the ultra-relativistic limit nor in the spherical body neither on the surface. Whether this solution is allowed physically or not is discussed in the discussion section, but to be brief it has not been proved mathematically that the spin zero limit of a stationary solution must also satisfy energy conditions; although it is greatly preferable by intuition.

\vspace{1cm}

\begin{tabular}{ l | ccccccccc  r }
  \hline
\rule{0pt}{3ex}   			
$ r_{g} \hspace{0.1cm} (m) / a \hspace{0.1cm}(m^2/s)$ & $1$ & $10$ & $10^2 $ & $10^3$ & $10^4 $ & $10^5 $ & $10^6 $ & $10^7 $ & $10^8 $  \\
\hline
\rule{0pt}{3ex}   
 $\hspace{0.9cm}10^{3}\hspace{0.1cm} (m)$ & V & V & V & S & S & S & S & S & S \\
$\hspace{1cm} 10^{4} \hspace{0.1cm} (m)$ & V & V & V & S & S & S & S & S & S  \\
 $\hspace{1cm}10^{5}\hspace{0.1cm} (m)$ & V & V & V & V & V & S & S & S & S\\
 $\hspace{1cm}10^{6}\hspace{0.1cm} (m)$ & V & V & V & V & V & V & S & S & S\\
 $\hspace{1cm}10^{7}\hspace{0.1cm} (m)$ & V & V & V & V & V & V & V & S & S\\
 $\hspace{1cm}10^{8}\hspace{0.1cm} (m)$ & V & V & V & V & V & V & V & V & V\\
\end{tabular}

 \vspace{ 0.5cm}
Table 1 : $a (m^2/s)$ versus $r_{g} (m)$ for an anisotropic source with $r_{s}=0.001  \hspace{0.1cm} N s^2/kg$ . S indicates that SEC, WEC, and DEC are satisfied while V is the sign of energy conditions' violation.

 \vspace{ 0.5cm}

\end{document}